\documentclass[12pt,preprint,a4paper]{aastex}
\usepackage{times}
\usepackage{graphics,epsf}
\usepackage{amsmath}                
\usepackage{amsfonts}               
\usepackage{amssymb}                
\usepackage{epsfig}                 
\usepackage{epstopdf}
\usepackage{rotating}
\usepackage{color}
\usepackage{tabularx}

\newcommand{\cm}{{~\rm cm}}
\newcommand{\s}{{~\rm s}}

\newcommand{\g}{{~\rm g}}

\newcommand{\yr}{{~\rm yr}}

\def \apj{ApJ}
\def \aap{A\&A}

\def \mnras{MNRAS}
\def \apjl{ApJ Lett.}

\def \nat{Nature}

\begin{document}

 \title{Binary interactions with high accretion rates onto main sequence stars}


 \author{Sagiv Shiber, Ron Schreier and Noam Soker\altaffilmark{1}}
 \altaffiltext{1}{Department of Physics, Technion -- Israel Institute of Technology, Haifa 32000 Israel; soker@physics.technion.ac.il.}

\begin{abstract}
Energetic outflows from main sequence stars accreting mass at very
high rates might account for the powering of some eruptive objects,
such as merging main sequence stars, major eruptions of luminous
blue variables, e.g., the Great Eruption of Eta Carinae, and other
intermediate luminosity optical transients (ILOTs; Red Novae; Red
Transients). These powerful outflows could potentially also supply the extra
energy required in the common envelope process and in the grazing
envelope evolution of binary systems.
{{{{We propose that a  massive outflow/jets mediated by magnetic fields might remove energy and angular momentum from the accretion disk to allow such high accretion rate flows.}}}} By examining the possible activity of the magnetic fields of accretion disks we conclude
that indeed main sequence stars might accrete mass at very high
rates, up to $\approx 10^{-2} M_\odot \yr^{-1}$ for solar type
stars, and up to $\approx 1 M_\odot \yr^{-1}$ for very massive
stars. We speculate that magnetic fields amplified in such extreme
conditions might lead to the formation of massive bipolar outflows
that can remove most of the disk's energy and angular momentum. It
is this energy and angular momentum removal that allows the very
high mass accretion rate on to main sequence stars.
\end{abstract}

{\bf Keywords: }
Stars: stars: AGB and post-AGB -- stars: jets -- (stars:) binaries (including multiple): close

\section{INTRODUCTION}
\label{sec:introduction}

Evidences are acquired in recent years on the need to allow
compact objects to accrete mass at rates exceeding the Eddington
limit by about one to two orders of magnitude. Extreme cases of
super-Eddington accretion rates is the magnetized neutron star
M82X-11 that accretes from a massive companion at a rate of
$\approx 100$ times the Eddington limit \citep{Bachettietal2014},
and two active galactic nuclei (AGNs) with $\approx 200-300$ times
the Eddington limit \citep{Duetal2015}. Other examples include
accretion onto a main sequence star during mergers of stars
\citep{SokerKashi2011}, {{{{during periastron passages of main sequence stars near asymptotic giant branch stars \citep{Staffetal2015},}}}}
and outbursts of some massive stars that might be power by accretion onto a main sequence star (e.g., \citealt{KashiSoker2010a, KashiSoker2010b, Kashietal2013}); a
general name for these type of objects is intermediate luminosity
optical transients (ILOTs; also termed Red Novae or Optical
Transients).  In many cases, super-Eddington or Sub-Eddington
outflow kinetic power is larger than the radiation luminosity,
e.g., the AGN HS~0810+2554 \citep{Chartasetal2014}.
\cite{Chartasetal2014} suggest that this implies that magnetic
activity is likely behind the acceleration of the outflow.

{{{{Despite the observational hints for high mass accretion rates onto main sequence stars, it is commonly assumed that main sequence stars cannot accrete mass at high rates, above $\approx 10^{-3} M_\odot \yr^{-1}$ (e.g., \citealt{HjellmingTaam1991}). Our main theme in the present study is to introduce the idea that main sequence stars might accrete mass at very high rates. This has implications to transient events, to the common envelope evolution, and to the newly proposed grazing envelope evolution. We approach this challenge by building on a strong magnetic activity in the accretion disk, including an efficient dynamo and magnetic field reconnection.}}}}
We deal only with main sequence stars, and with accretion rates of up to
$\approx 0.01 M_\odot \yr^{-1}$ for low mass main sequence stars,
and up to $\approx 1 M_\odot \yr^{-1}$ for very high mass main
sequence stars. In the merger model for the ILOT V838~Mon
\citep{SokerTylenda2003}, for example, a companion of $\approx 0.3
M_\odot$ was destroyed on a star of $\approx 6 M_\odot$. It is
assumed that several percents of the companion mass was accreted
within several months; the rest inflated a large envelope. This
leads to an accretion rate of $\approx 0.01-0.1 M_\odot \yr^{-1}$.
{{{{We note that \cite{Nandezetal2014} simulate a merger process with parameters appropriate for the ILOT V1309~Sco, and find the formation of an accretion disk in the synchronized case. They did not find jets, as expected if magnetic fields are not included.
As well, in the non-synchronized case the merger proceeds without the formation of an accretion disk. Namely, it is possible, {{{{{ according to their results, }}}}} that merger and high accretion rate does not necessary requires an accretion disk.}}}}

It is commonly agreed upon that jets are produced at the center of
accretion disks and that magnetic fields are a crucial ingredient
in their launching process (e.g., \citealt{Livio2011,
Pudritzetal2012}). Many models for the formation of astrophysical
massive outflows are based on the operation of large scale
magnetic fields within and around the accretion disk (e.g.,
\citealt{ZanniFerreira2013} and \citealt{Narayanetal2014} and
references therein).  By large-scale we refer to magnetic fields
with coherence scale larger than the radius of the disk at the
considered location.  In the "X-wind mechanism" introduced by
\cite{Shuetal1988} and \cite{Shuetal1991} the jets are launched
from a narrow region in the magnetopause of the stellar field. A
different group of models also use open radial magnetic field
lines and are based on an outflow from an extended disk region,
but they do not rely on the stellar magnetic field (e.g.,
\citealt{KoniglPudritz2000, Shuetal2000, Ferreira2002,
Krasnopolskyetal2003, FerreiraCasse2004}).

There are several arguments that point to problems with models for
launching jets that are based only on large scale ordered magnetic
fields. (1) Precessing jets. In several young stellar objects
(YSOs) the jets precess on a time scale of 100 years (e.g.
\citealt{Ybarraetal2006}). A large scale magnetic field cannot
change its symmetry axis on such a short time. (2) Collimated jets
in planetary nebulae (PNe). There is a highly collimated clumpy
double-jet in the PN Hen~2-90 \citep{SahaiNyman2000} very similar
in properties to jets from YSOs that form Herbig-Haro (HH) objects
(dense blobs along jets launched by YSOs). The source of the
accreted mass in PNe is thought to be a companion star. In such a
system large scale magnetic fields are not expected. (3) No jets
in DQ Her (intermediate polars) systems. DQ Her systems are
cataclysmic variables where the magnetic field of the accreting
white dwarf (WD) is thought to truncate the accretion disk in its
inner boundary. This magnetic field geometry is the basis for some
jet-launching models in YSOs (e.g., \citealt{Shuetal1991}).
However, no jets are observed in intermediate polars. (4) Some of
these models don't take into account the strong dynamo that is
expected to operate in accretion disks. In few cases, e.g.,
\cite{Hujeiratetal2003}, the modification of the large scale
magnetic field in the disk is considered.  Such dynamos
substantially modify the structure of the magnetic field in the
accretion disk and lead to a behavior closer to that in active
stars. (5) \cite{Ferreira2013} summarizes accretion models of
classical T Tauri (TTS) stars, and concludes that steady state
models, such as X-winds or Accretion Powered Stellar Winds, cannot
spin down a protostar. The balance between the accretion torque
(working so as to spin up the star) and the magnetic torque
(working so as to slow the stars' rotation) should lead to a spin
up on a time scale smaller than the disk lifetime, whereas
observations show that the rotation period of TTSs is almost
constant, at roughly $10 \%$ of their break-up speed. Instead,
unsteady models, such as Magnetospheric Ejection are the best
candidates for removal of extra energy from the disk. The model
suggested in this study is a type of unsteady model.

{{{{{ In the present paper we take the view that }}}}} the very high accretion rates onto main sequence stars in ILOTs,
in the range of $\approx 10^{-4}-1 M_\odot \yr^{-1}$
\citep{KashiSoker2010b}, require the formation of thick accretion
disks.
{{{{The accretion rates in our model are high,
and the disks are thick enough so as to sustain turbulent eddies
for long durations. Those eddies, combined with the
differential rotation of the disk's plasma, generate
the dynamo magnetic fields. Strong fields will eject
mass and momentum out of the disk without the need of
inward diffusion.}}}}
In section \ref{sec:disk} we build such accretion disks
that we will use in section \ref{sec:magnetic} where we study
their magnetic field properties and activity. We will take the
magnetic dynamo to be similar in many respects to that of a
stellar dynamo. The merit of comparison to the dynamo operating in
the Sun is demonstrated in a recent paper by \cite{Buglietal2014}.
They study dynamo inside a thick disk around a Kerr black hole and
find the evolution of the magnetic field to qualitatively occur in
the same fashion as in the Sun. Our summary is in section
\ref{sec:summary}.

\section{BUILDING THE DISK}
\label{sec:disk}

\subsection{A thick disk}
\label{subsec:thickDisk}

{{{{We here construct an accretion disk with a high accretion rate where part of the energy is carried away by jets. High accretion rate disks were studied before, e.g., \cite{Abramowiczetal1980} for an accreting BH, as well as accretion disks where heat is removed to heat the corona (e.g., \citealt{SvenssonZdziarski1994}), or to an outflow (e.g., \citealt{KuncicBicknell2006}). To present a comprehensive picture we repeat the construction of the disk presented in these studies, and present it in a way that will serve our goals.}}}}

{{{{
\cite{SvenssonZdziarski1994} modified the standard Shakura-Sunyaev model for accretion disks
\citep{ShakuraSunyaev1973, Franketal2002} by allowing a major fraction, $f$, of the power released
from the accreted disk matter to be transported and dissipated in the corona.
We replace their $(1 - f)$ factor by}}}}
the parameter $\varepsilon$ that is the fraction of the energy
liberated by the accretion process that goes to radiation
\begin{equation}
\frac{4\sigma T_{c}^{4}}{3\tau}  =
\varepsilon\frac{3GM\dot{M}}{8\pi
R^{3}}\left[1-\left(\frac{R_{*}}{R}\right)^{\frac{1}{2}}\right].
 \label{eq:EnergyEquation}
\end{equation}
In our study the rest of the energy is channelled to the
outflowing gas, while the contributions to the pressure from this
channels, e.g., magnetic fields, are neglected. The other symbols
have their usual meaning: $M$ is the mass of the star, $\dot M$ is
the mass accretion rate, $R$ is the radial coordinate in the disk
plane, $R_\ast$ is the stellar radius, $\tau$ is the optical
depth, and $\sigma$ is Stefan-Boltzmann constant. We study
accretion disks with very high mass accretion rates of $\approx
10^{-4}-1 M_\odot \yr^{-1}$ in which the primary star is a
main-sequence star with a radius of $R_\ast \approx 10^{11} \cm$.
This can cause the opacity to be dominated by electron scattering
processes rather than Kramers opacity induced by free-free
emission, and also the pressure to be dominated by radiation
pressure rather than gas pressure.

In the gas pressure-dominated regime we take only gas pressure,
starting with the Kramers opacity. We will use a molecular weight
of $\mu=0.615$ throughout the paper. When the parameter
$\varepsilon$ is included, we find the following expressions for
the
{{{{radial dependance of the}}}}
disk height $H$, the central temperature $T_c$, and the radial
velocity of the disk $v_R$, respectively
\begin{equation}
H/R  =  0.45\varepsilon^{1/20}\alpha^{-1/10}m^{-3/8}R_{12}^{1/8}\left(f^{4}\dot{M}_{24}\right)^{3/20},
 \label{eq:gasWithKramersResultHeight}
\end{equation}
\begin{equation}
T_{c}  =  2\times10^{5}\varepsilon^{1/10}\alpha^{-1/5}\left(\frac{m}{R_{12}^{3}}\right)^{1/4}\left(f^{4}\dot{M}_{24}\right)^{3/10}\,{\rm K},
 \label{eq:gasWithKramersResultTemp}
\end{equation}
and
\begin{equation}
v_{R}  =  -3.5\times10^{6}\varepsilon^{1/10}\alpha^{4/5}m^{-1/4}R_{12}^{-1/4}\dot{M}_{24}^{3/10}f^{-14/5}\,{\rm cm\, s^{-1}},
 \label{eq:gasWithKramersResultVelocity}
\end{equation}
where $\alpha$ is the viscosity disk parameter which assumed to be
radius-independent, and we defined the  dimensionless variables
$$ m \equiv \frac{M}{M_{\odot}},
\;\;\;R_{12} \equiv \frac{R}{10^{12}\cm},
$$
\begin{equation}
\dot{M}_{24} \equiv \frac{\dot{M}}{10^{24}\g\s^{-1}},
\;\;\;f^4 \equiv 1-\sqrt{\frac{R_{*}}{R}}.
\label{eq:fm24Definition}
\end{equation}
For these typical values the disk is no longer thin, and the
radial velocity has the same order of magnitude as the Keplerian
velocity. In addition, the high temperature suggests that electron
scattering processes and radiation pressure might be important.

Taking the pressure to be radiation pressure and taking electron
scattering opacity give the following disk properties
\begin{equation}
H/R  =  1.65\varepsilon\left(f^{4}\dot{M}_{24}\right)R_{12}^{-1},
 \label{eq:RadsWithESResultHeight}
\end{equation}
\begin{equation}
T_{c}  =  1.3\times10^{5}\varepsilon^{-1/4}\alpha^{-1/4}\left(\frac{m}{R_{12}^{3}}\right)^{1/8}\,{\rm K},
 \label{eq:RadsWithESResultTemp}
\end{equation}
and
\begin{equation}
v_{R}  =  -4.65\times10^{7}\varepsilon^{2}\alpha m^{1/2}R_{12}^{-5/2}\dot{M}_{24}^{2}f^{4}\,{\rm cm\, s^{-1}}.
 \label{eq:RadsWithESResultVelocity}
\end{equation}
The properties are more sensitive to the parameter $\varepsilon$,
which can strongly affect the thickness and the radial velocity of
the disk. {{{{As noted in previous works (e.g., \citealt{SocratesDavis2006}),}}}}
it is possible to build thin disks even for
very high accretion rates by taking $\varepsilon \ll 1$. Another
important difference is that the disk height over radius is
a decreasing function with radius where in the gas pressure
dominated case it is an increasing function with radius.

{{{{{ In any case, in order not to required extremely small value of $\varepsilon$, and demanding that $H$ is not much larger than $r$, we find from the expressions for $H/r$ that the accretion rate in our treatment is limited. To an order of magnitude this limitation on the accretion rate is  $\approx 10^{-2} M_\odot \yr^{-1}$ for solar type stars, and up to $\approx 1 M_\odot \yr^{-1}$ for stars of 30 and more solar masses. }}}}}

\subsection{Relevant time scales}
\label{subsec:timescales}

There are several relevant time scales. The dynamical timescale
(or the Keplerian time divided by $2 \pi$) is given by
\begin{equation}
t_{\phi} \approx \frac{R}{v_{\phi}} \approx \Omega_{{\rm K}}^{-1}.
 \label{eq:kplerianTimeScale}
\end{equation}
This time scale is about equal to the sound crossing time along
the disk height, which is about the dynamical time scale to reach
equilibrium along the $z$ direction (perpendicular to the disk
plane)
\begin{equation}
t_{z} \approx \frac{H}{c_{s}},
 \label{eq:ZTimeScale}
\end{equation}
where $c_s$ is the sound speed.
 The viscous time scale is given by
\begin{equation}
t_{{\rm visc}} \approx \frac{R^{2}}{\nu} \approx \frac{R}{v_{R}},
 \label{eq:ViscousTimeScale}
\end{equation}
where $\nu = \alpha c_s H$ is the turbulence viscosity
coefficient. The thermal timescale is given by
\begin{equation}
t_{{\rm th}}={\rm \frac{heat\, content\, per\, unit\,
area}{dissipation\, rate\, per\, unit\, area}} \approx
\frac{\Sigma c_{s}^{2}}{D\left(R\right)}.
 \label{eq:ThermalTimeScale}
\end{equation}
Where $D\left(R\right)$ is the dissipation due to differential
viscous torque and $\Sigma$ is the surface density of the disk.

In the Radiation dominated regime it can be shown that
\begin{equation}
t_{{\rm th}}=\varepsilon\frac{H}{c}\tau=\varepsilon \frac{t_{\rm{diff}}}{3},
 \label{eq:ThermalDiffusionRelations}
\end{equation}
where $t_{\rm{diff}}=3H\tau/c$ is the photon diffusion time from
the disk's mid-plane out. From equation
(\ref{eq:ThermalDiffusionRelations}) we can interpret
$\varepsilon$ as the probability to lose energy (heat) via photon
diffusion (hence radiation) rather than outflow kinetic energy.
Approximately, as the diffusion time becomes longer a larger
fraction of the energy is transported by other than radiation
channels. In our model these are magnetic fields and gas outflows.
{{{{We note that when magnetic fields remove energy much faster than photons do, then the disk might be geometrically thin, but still the radiation carries only a small fraction of the energy.}}}}

Let us dwell on this point. The viscosity in the disk and magnetic
activity dissipate energy. But in our regime of very high
accretion rate, the photon diffusion time is very long. Most of
the dissipated energy that starts as thermal energy, rather than
being emitted by the disk, will lead to internal motion, such as
turbulence that amplifies magnetic fields, and to gas buoyancy
outward. Near the surface, in particular due to magnetic field
reconnection below the photosphere (see section
\ref{sec:magnetic}), the energy is channelled to winds or jets.
The photons have no time to diffuse, and most of the energy is carried by
the outflow, i.e., kinetic energy.
{{{{The acceleration occurs in many sporadic
impulsive events resulting from magnetic field reconnection.
Most of the reconnection events are much shorter than the dynamical time,
and hence the mass and energy they eject merge to one continues outflow.}}}}

For the treatment above to hold, the inflow time must be longer
than one rotation time of material in the disk
\begin{equation}
t_{{\rm in}} \ga  t_{{\rm kep}} = 2 \pi t_{\phi}.
 \label{eq:shortKeplerianTime}
\end{equation}
This condition can be cast to height over radius condition. With
the timescales definitions $t_{\rm in}= R/v_R$, $t_{\phi}=H/c_s$
and using the relation $ v_R=\left( 3 \nu / 2R \right) f^{-4}$ one
gets
\begin{equation}
\frac{1}{2\pi} \ga \alpha\frac{3}{2}\left(\frac{H}{R}\right)^{2}f^{-4}.
\label{eq:shortKeplerianTimeHeightOnRadius}
\end{equation}
The area in the $\varepsilon-\dot{M}$ plane where this inequality
holds is termed the `safe-zone'. For others values of $\alpha$,
$R_\ast$, and $M$ the safe-zone changes.

\section{MAGNETIC ACTIVITY}
\label{sec:magnetic}

\subsection{Dynamo}
 \label{subsec:dynamo}

The high accretion rate is associated with very large Reynolds
flows and therefore the transition to turbulence  is inevitable.
The differential rotation in the disk is responsible for the
generation of the toroidal magnetic field, the $\Omega$-effect,
and the combined action of helical turbulence and differential
rotation generates the poloidal field, the $\alpha$-effect, and so
forth. This is the basic of the $\alpha$-$\Omega$ dynamo. We
follow the treatment of the magnetic field amplification as
conducted by \cite{Pudritz1981a}.

We study flow regimes with very high Reynolds numbers.
\cite{NathMukhopadhyay2015} have shown recently that for high
Reynolds numbers, as is the case for realistic astrophysical
accretion disks, nonlinearity is achieved by transient growth
(e.g., \citealt{UmurhanRegev2004, Umurhanetal2007}) rather by the
Magnetorotational Instability (MRI). \cite{NathMukhopadhyay2015}
also showed that when the transient growth modes get to play, the
magnetic fields increase much more than what the pure MRI model
gives. For that, although the results obtained by
\cite{Pudritz1981a} might not be applicable when MRI dominates, in
the regime studied here the transient growth dominates and we use
the results by \cite{Pudritz1981a}.

 The dynamo action is described by the mean
field theory, in which the magnetic field and the velocity are
decomposed to mean and fluctuating components,
\begin{equation}
\vec{B_t}=\vec{B} + \vec{b}, \qquad \vec{U_t}=\vec{U} + \vec{u}.
\label{eq:decomp}
\end{equation}
The induction equation for the fluctuating component of the
magnetic field is written under the so called `first order
smoothing theory' as
\begin{equation}
\frac {\partial \vec{b}}{\partial t} =
 \vec{\nabla} \times \left( \vec{U} \times \vec{b} + \vec{u} \times \vec{B} \right) + \eta {\nabla}^2 \vec{b} .
\label{eq:induct1}
\end{equation}
The last term comes from Ohmic dissipation of the fluctuating
component of the current, given by $\vec{j}=(c/4 \pi) \nabla
\times \vec{b}$, and reduces the magnetic field. The first term on
the right hand side might lead to convective amplification of the
magnetic field, the `$\alpha$-dynamo'. From the assumption that
the Ohmic dissipation rate of fluctuating magnetic energy per unit
volume $\dot e_{\rm Oh} =(\eta/ l_u^2)b^2/4 \pi$, equals the rate
at which turbulence pump energy back $\dot e_{\rm tr} = (\eta_T/
l_u^2) B^2/4 \pi$, where $l_u$ is the largest turbulence scale,
\cite{Pudritz1981a} derived the magnitude of the fluctuating
magnetic field
\begin{equation}
{b}^2 =
    \frac {\eta_T}{\eta} B^2.
\label{eq:b2}
\end{equation}
The magnetic diffusivity $\eta$ is given by the expression
\begin{equation}
\eta \equiv c^2/4\pi\sigma_s =
2.95\times 10^{11} \ln\Lambda\  T^{-3/2}\ \ {\rm cm^2/s},
\label{eq:eta}
\end{equation}
where $\sigma_s$ is the Spitzer conductivity, $T$ is the temperature,
and $\ln\Lambda$ is of order 10 (see \citealt{Huba2013}).
$\eta_T$ is the turbulent diffusivity for the mean
magnetic field $B$,
\begin{equation}
\eta_T = M^2_t H^2/t_K.
\label{eq:etaT}
\end{equation}
Here $H$ is disk height, $t_K$ the Keplerian orbital
period, and $M_t$ is the turbulent Mach number defined as
\begin{equation}
M_t \equiv {u}/c_s,
\label{eq:Mt}
\end{equation}
with ${u}$ the r.m.s. turbulent speed, and $c_s$ is the speed of sound.

Substituting equation (\ref{eq:gasWithKramersResultHeight}) into (\ref{eq:etaT}),
and (\ref{eq:gasWithKramersResultTemp}) into (\ref{eq:eta})
we find the ratio
\begin{equation}
\begin{split}
{b}^2 /B^2 = \eta_T/\eta
= 3.2\times10^{13}
\left(\frac{M_t}{0.5}\right)^2
\left(\frac{\varepsilon}{0.1}\right)^{1/4}\\
\times\left(\frac{\alpha}{0.1}\right)^{-1/2}
\left(\frac{m}{R_{12}^{3}}\right)^{1/8}
\left(f^{4}\dot{M}_{24}\right)^{3/4}.
\end{split}
\label{eq:b_over_B}
\end{equation}

There are several large uncertainties in the derivation of
equation (\ref{eq:b_over_B}). Nonetheless, its implication is
clear: the fluctuating magnetic field dominates. Any initial large
scale, of the order of the disk radius and more, magnetic field
becomes negligible in the disk. At these high accretion rates
large scale magnetic fields are not expected to play a significant
role in the ejection of jets or disk winds. We should look for the
role played by the fluctuating magnetic field, that is amplified
by the dynamo, in launching the jets and winds.

The ratio of the fluctuating magnetic pressure to the total
pressure is defined by
\begin{equation}
\lambda \equiv {b}^2 /4\pi P,
 \label{eq:lambdad}
\end{equation}
so that the mean magnetic field is
\begin{equation}
B =\lambda^{1/2} \sqrt{4\pi P\ \eta/\eta_T}.
\end{equation}
Taking $P=\rho c_s^2$, \cite{Pudritz1982} estimated a very large
value $\lambda$ of
\begin{equation}
10^{7/2} \leq \lambda \leq 10^{4}.
 \label{eq:lambda}
\end{equation}
As we are discussing accretion into main sequence stars where we
apply the disk structure neglecting magnetic pressure (section
\ref{sec:disk}), we limit the ratio of the magnetic to gas
pressure to $\lambda \la 1$.  As well, we take a conservative
approach in this preliminary study.  We derive the following value
for the magnetic field
\begin{equation}
\begin{split}
|b| = & 2 \times10^4 \sqrt{\lambda}
\left(\frac{\varepsilon}{0.1}\right)^{-1/40}
\left(\frac{\alpha}{0.1}\right)^{-9/20} \\ &\times \left(
\frac{m}{R_{12}^3} \right)^{7/16} \left(f^4
\dot{M}_{24}\right)^{17/40} \,\rm{G}.
\end{split}
\label{eq:abs_b}
\end{equation}

{{{{Two comments are in place here regarding the numerical value given in equation (\ref{eq:abs_b}). ($i$) In solar spots the magnetic fields reach equipartition, between the magnetic and thermal pressure, for magnetic fields of $\approx 2000 \,\rm{G}$. The magnetic fields in the accretion disk studies here are about one order of magnitude stronger. This implies that the magnetic activity in the accretion disk is extremely strong, with magnetic pressures about two orders of magnitude above the photospheric pressure of the accreting star.
{{{{{ It should be noted that for the very high accretion rates used here $\ga 10^{-2} M_\odot \yr^{-1}$ and the low value of $\varepsilon \approx 0.1$, both the density and temperature inside the disk are much higher than the corresponding values on the stellar photosphere. The pressure inside the disk is therefore higher by more than two orders of magnitude than that on the stellar surface, and the strong magnetic field is still below the thermal pressure inside the disk. }}}}}
($ii$) The scaling in equation (\ref{eq:abs_b}) takes $\lambda=1$. This is applicable to the case where the magnetic fields reaches equipartition with the thermal pressure in the accretion disk, as is the situation in solar spots. Above the solar surface the magnetic pressure can be much larger than the thermal pressure. Similarly, it is quite possible that in the accretion disk we study here the magnetic pressure will be larger than the thermal pressure, and rather be equal to the ram pressure of the Keplerian motion in the disk  $\rho v_\phi ^2$. This is the case for $\lambda > 1$, as taken by \cite{Pudritz1982}. Such magnetic fields will make the proposed scenario more feasible even.}}}}

\subsection{Magnetic power}
 \label{subsec:power}

We show now that most of the energy could potentially be carried off by magnetic
activity rather than by radiation even when the magnetic energy in
the disk is smaller than the thermal energy.

{{{{We assume that most of the magnetic activity takes place just inward to a radius $R_B$, not too far from the stellar surface. We neglect further amplification of the magnetic field inward to this radius. }}}}
The mass accretion rate is given by $\dot M \approx 4 \pi R_B H \rho
v_R$. We define the ratio of the inward radial velocity to the
Keplerian velocity $\delta =v_R / v_\phi <1$, and the kinetic
energy density in the disk $e_d \simeq \rho v_\phi ^2/2$. The
accretion power {{{{ of material crossing the radius $R_B$ inward, i.e., the energy flux through radius $R_B$,}}}} is given by
\begin{equation}
L_{\rm acc}= \frac{1}{2}\dot M v_\phi ^2 \approx 2 \pi \delta H R_B
v_\phi e_d.
 \label{eq:lacc}
\end{equation}
We take the time to amplify and release magnetic energy to be the
dynamical time given by equation (\ref{eq:ZTimeScale}) $\tau_B
\approx t_{z} \approx {H}/{c_{s}} \approx R_B/v_\phi$. With an
energy density $e_B= b^2 / 8 \pi$ the magnetic power
{{{{within radius $R_B$}}}} is $L_B \approx 2 \pi R_B^2 H e_B/ \tau_B$.
 The ratio of magnetic energy density to kinetic energy density is then
\begin{equation}
\frac{e_B}{e_d} \approx  \delta \frac{L_B}{L_{\rm acc}}.
 \label{eq:eb}
\end{equation}
{{{{We recall that in deriving equation  (\ref{eq:eb}) we assumed that there is a global steady state, and hence the amplification time of the magnetic field is equal to the time scale to remove magnetic energy by reconnection. Hence, for a given power, shorter amplification and reconnection time scales imply weaker magnetic fields. Alternatively, if the ratio of the magnetic energy density to the kinetic energy density is some fixed value, like unity, then the magnetic power increases with decreasing time scales.}}}}

As $\delta <1$, the above simple estimate shows that the magnetic
energy can be lower than the kinetic energy in the disk, and yet
the magnetic activity might carry a large fraction of the accretion
energy. For that the magnetic energy should be vigorously
amplified by a vigourous turbulence in the disk. We assume that
the high accretion rate leads to this behavior of the disk.

\subsection{Reconnection and outflow}
 \label{subsec:reconnection}

{{{{In this subsection we speculate that the magnetic activity might lead to an energetic outflow.}}}}

We define three general coherence scales of the magnetic fields.
Small scale defined as much smaller than the disk scale hight
$\l_B \ll H$. Moderate scale for which the coherence scale is of
the order, but smaller, than the disk scale hight, $0.3 H \la l_B
< H$. Finally, the large scale fields is with $l_B \ga r
> H$. In the present study  magnetic fields with moderate scale, {{{{about equals to the turbulence scale,}}}} dominate. We give no role to large scale magnetic fields (unlike many other jets' models).

As magnetic flux accumulated and the tension increases due to the
dynamo action, moderate-scale turbulence trigger local explosive
events of magnetic reconnection \citep{Lazarianetal2015}. Those
events are sporadic and decay as the extra magnetic energy is
released. At any given radius $R$ the largest turbulence scale,
$l_u$, is given by $l_u=M_t\cdot H$ \citep{Pudritz1981a}, and for
high turbulent Mach numbers such as $M_t\sim0.5$ expected for the
presently studied accretion flows (section \ref{sec:disk}),  the
vertical scale can be populated by only 2 such large eddies. The
large magnetic flux tube will also buoy quite fast, relative to
small flux tubes, outward from the disk, allowing the replenishing
time of the magnetic fields to be of the order of the dynamical
time.

 There are two key processes resulting from the large flux tubes.
These are magnetic torque and reconnection in scales not much
smaller than the disk scale height. In accretion disk models where
the large scale magnetic fields extended beyond the disk, there is
a magnetic torque on the disk from gas outside the disk, mainly a
disk wind. When there is no such a field, torque acts within the
disk. When the flux tubes are small relative to the radius, the
torque is local and play no large scale role. In our situation the
flux tubes are within the disk, but they are very large.
Therefore, {{{{we speculate that }}}} they might transfer angular momentum and energy from the internal parts of the disk to near the surface of the disk. Near the surface magnetic field
{{{{are likely to}}}} reconnect and accelerates the wind/jets. This process further reduces the amount of energy that is radiated from the disk, hence lowering the value of
$\varepsilon$.

The vigourous turbulence facilitates magnetic field reconnection \citep{Lazarianetal2015}. {{{{We propose}}}} therefore, that magnetic field reconnection will take place not only above the disk, as in the Sun, but also below the disk surface. The outcomes of the reconnection process of the large flux tubes {{{{might be}}}} as follows.

(1) \emph{Mass outflow.} Like in the Sun, once reconnection occurs
between two adjacent magnetic flux tubes plasma flows along
reconnected magnetic field lines, in two opposite directions.  In
our setting {{{{we take that}}}} a large fraction of the reconnection events takes place below the photosphere. Although initially a large fraction of the dissipated energy is thermal, before photons have time to diffuse out the gas is accelerated to form an outflow. Each
reconnection event is like a small explosion. Much as in supernova explosions where photons diffusion time is very long and most of the final energy is kinetic with only a small fraction emitted as radiation.

 (2) \emph{Huge magnetic arcs.} The reconnection process implies more than direct mass ejection, and migth lead to the build up of huge magnetic arcs above the disk. While the original flux tube
emerging from the disk are expect to reside at about the same
radial distance $R$, {{{{we speculate that}}}} the new arcs formed by reconnection can have their two foot-points at different radii, much as trans-equatorial
magnetic loops in the Sun \citep{Tadesseetal2014}. We term these {{{{speculative structures}}}} trans-radial loops. The trans-radial loops experience very large shear, and might lead to further amplification of the magnetic energy. This process will be studied in the next paper in the series   (see also \citealt{HeyvaertsPriest1989}).

 (3) \emph{Flares.} Reconnection of such huge arcs might lead to extremely energetic
flares. Such an energetic flare was speculated to be the cause of
the 2009 November outburst of Aquila X-1 \citep{Soker2010}.
\cite{DalPino2010}, \cite{Khialietal2015} and
\cite{Kadowakietal2015} discuss such rapid reconnection processes
in relation to different properties of accretion disks of
microquasars. We aim here at accretion disks around main sequence
stars accreting at a very high rate. We expect the wind from the
disk to obscure radiation from such flares.

\section{SUMMARY}
 \label{sec:summary}

We described a crude accretion flow setting to allow main sequence
stars to accrete mass at very high rates. The key to allow such
high accretion rates is that most of the energy liberated in the
accretion process, and the angular momentum carried by the
inflowing gas, are removed by massive outflows.
{{{{\cite{SvenssonZdziarski1994} used the removed energy to heat the corona, and \cite{KuncicBicknell2006} produced jets by magnetic torque, both in accretion disks onto black holes. We here studied accretion onto main sequence stars, and the removal of energy and angular momentum by jets that are formed from sporadic reconnection events of magnetic fields that are amplified by a strong dynamo. While in previous papers of our group it was just assumed that main sequence stars might accrete at very high rate, in the present study we have outlined the scenario for that. The main result is that the removal of angular momentum and energy cannot be achieved without magnetic fields that are efficiently amplified by the operation of a dynamo in the dense accretion disk.}}}}

{{{{As was done in the past (e.g., \citealt{SvenssonZdziarski1994})}}}} we described a disk structure where only a fraction
$\varepsilon \ll 1$ of the energy liberated by accretion is
channelled to radiation (section \ref{sec:disk}). From this crude
structure we could constrain the properties of the disk around
main sequence stars. For the parameters used here we found a thick
accretion disk dominated by thermal pressure (eqs.
\ref{eq:gasWithKramersResultHeight} -
\ref{eq:gasWithKramersResultVelocity}). To still be regarded as an
accretion disk in the usual meaning, the inflow time should be
longer than the orbital time.

Using the disk properties, we {{{{speculated on}}}}  possible magnetic
activities that might lead to rapid mass ejection from the disk. We
presented the expected dynamo amplification of the magnetic field,
and concluded that a very strong fluctuating magnetic field is
expected in the disk (eq. \ref{eq:b_over_B}). We then argued
(section \ref{subsec:power}) that it is possible for the magnetic
power to be very large without altering the general disk
structure. We also note that for geometrically thin disks, for
which $\delta \ll 1$, the removal of energy by magnetic fields can
be efficient even for $e_B \ll e_d$. Such a thin disk might be
obtained if the disk is radiation dominated, as seen from equation
(\ref{eq:RadsWithESResultHeight}) for $\varepsilon \ll 1$.
{{{{This outlined scenario for allowing high mass accretion rates onto main sequence stars, although speculative in some parts, is our main new result. We hope it will encourage relevant models of astrophysical systems, such as the common envelope evolution, to consider main sequence stars that accrete mass at very high rates, as well as motivate numerical studies of the accretion flow. Such 3-dimensional numerical studies should include magnetic fields and be of high spatial resolution to resolve the turbulence in the accretion disk.}}}}

Some key ingredients of the proposed scenario should be emphasized.
(1) In the proposed model magnetic fields with moderate coherence
scales dominate. By moderate scale we refer to scale of $0.3 H \la
l_B <H$, where $H$ is the disk scale-height. Large-scale magnetic
fields of $l_B \gg H$ play no role in our model (unlike in many
other jets' models).
 (2) Although magnetic energy is dissipated locally to heat by
reconnection, in our setting most of this energy ends up as
kinetic energy of the outflow. The reason is that because of the
very high mass accretion and outflow rates, the flow is optically
thick. Therefore, before photons have time to diffuse out the gas
expands. Most of the thermal energy goes to adiabatic expansion
rather than to radiation (much as in supernovae explosions).

The outcomes of strong magnetic field reconnection below the
photosphere, discussed in section \ref{subsec:reconnection},
brought us to propose the existence of trans-radial magnetic field
lines,  similar to those studied by \citep{HeyvaertsPriest1989}.
 Namely, huge magnetic arcs above the disk that have their two
foot-points at different radii. These trans-radial magnetic fields
might lead to further magnetic field amplification. We expect
trans-radial loops to play a major role in the magnetic activity
of the boundary layer, where the disk angular velocity
substantially decreases to match the angular velocity of the star.
The dynamics of trans-radial loops is the subject of a forthcoming
paper.

Our main conclusion is that main sequence stars might accrete mass
at very high rates, up to $\approx 10^{-2} M_\odot \yr^{-1}$ for
solar type stars, and up to $\approx 1 M_\odot \yr^{-1}$ for very massive stars. Such accretion rates could potentialy account for some outbursting objects that have their luminosity in the
bulk region between novae and supernovae and that are thought to
be powered by high accretion rates onto main sequence stars in
binary systems \citep{KashiSoker2010b, KashiSoker2010b,
Kashietal2013}. As well, such high accretion rates might take place
in main sequence companions in eccentric orbits around AGB stars
\citep{Staffetal2015}. A general name for these types of objects is
intermediate luminosity optical transients (ILOTs; also termed Red
Novae or Optical Transients). Examples of specific objects include
stellar merger, such as V838~Mon \citep{{SokerTylenda2003}}, the
nineteenth century Great Eruption of Eta Carinae
\citep{KashiSoker2010a}, progenitors of some PNs
\citep{SokerKashi2012}, such as KjPn~8 \citep{BoumisMeaburn2013},
and the pre-explosion outbursts of SN~2009ip
\citep{SokerKashi2013}.

High accretion rates onto, hence high outflow rates from, main
sequence stars might supply extra energy to remove a common envelope
where the secondary is a main sequence star. When the envelope
removal by jets is sufficiently efficient as to remove the entire
envelope residing outside the secondary star, no real common
envelope phase is established. Instead, a grazing envelope
evolution (GEE) commences \citep{Soker2015}. The GEE might nicely
explain the observation that the kinetic energy of the bipolar
outflow from the binary system HD~101584 exceeds the released
orbital energy \citep{Olofssonetal2015}.

We can summarize our study by stating that the possibility for
main sequence stars to eject winds/jets at extremely high rates,
as we argued here, opens a rich variety of processes that could potentialy
account for some properties and some evolutionary puzzles of
binary stellar systems.

{{{{We thank an anonymous referee for very detail report and many valuable suggestions.}}}}

{}

\end{document}